# Least-squares fitting approach using energy, gradient and Hessian data to obtain an accurate quartic force field : Application to $H_2O$ and $H_2CO$.

Philippe CARBONNIERE*, Didier BEGUE, Alain DARGELOS and Claude POUCHAN

Laboratoire de Chimie Structurale – UMR 5624

Université de Pau et des Pays de l'Adour

IFR – Rue Jules Ferry

64000 PAU

Fax : 33 (0)5 59 80 37 69

*Email : philippe.carbonniere@univ-pau.fr

In this work we present an attractive least-squares fitting procedure to obtain a quartic force field by using energy, gradient and Hessian data arising from electronic wave function calculations on a suitably chosen grid of points. We use the experimental design to select the grid points : a "simplex-sum" of Box and Behnken grid is used for its efficiency and accuracy. We illustrate the numerical implementation of the method by using energy and gradient data and we test for $H_2O$ and $H_2CO$ the B3LYP/cc-pVTZ quartic force field performed from 11 and 44 simplex-sum configurations. Results compared to classical 44 and 168 energy calculations, show excellent agreement.

# I. INTRODUCTION

Calculation of molecular vibrations requires the construction of a potentiel function which can be written as a Taylor expansion in terms of curvilinear displacement coordinates :

$$V = V_{eq} + \sum_i \left(\frac{\partial V}{\partial s_i}\right)_{eq} s_i + \frac{1}{2!}\sum_{i\le j}\left(\frac{\partial^2 V}{\partial s_i \partial s_j}\right)_{eq} s_i s_j + \frac{1}{3!}\sum_{i\le j\le k}\left(\frac{\partial^3 V}{\partial s_i \partial s_j \partial s_k}\right)_{eq} s_i s_j s_k + \frac{1}{4!}\sum_{i\le j\le k\le l}\left(\frac{\partial^4 V}{\partial s_i \partial s_j \partial s_k \partial s_l}\right)_{eq} s_i s_j s_k s_l \ldots \quad (1)$$

Quadratic, cubic and quartic force constants are generally obtained by fitting data stemming from *ab-initio* calculations of the electronic energy for several nuclear configurations, or by a finite difference procedure of first or second derivatives of electronic energy with respect to the nuclear coordinates. Whatever the manner of proceeding, one to accumulates a great number of data resulting from *ab-initio* calculations carried out in a grid of points representing the geometrical variations and deduces from it the analytical potential function.

The number of force constants increases quickly with the size of the molecule, leading to problems of data acquisition and precision on the polynomial expansion coefficients to be determined for systems over four atoms.

Attempts to overcome this difficulty in the literature aim at obtaining quartic force fields by using energy, gradient and Hessian data arising from electronic wave function calculations[1,2,3,4], raising the question of the most efficient and accurate point distribution for the determination of polynomial parameters. These requirements are often neglected in papers relating to this problem, so we particularly lay emphasis on this aspect here.

In this work we present the procedure implemented in our code REGRESS EGH[5], for determining the analytical form of the potential with a reduced number of points to be calculated without deteriorating the accuracy of the results. These procedures include the “extended least squares fitting”, which consists jointly in fitting all the data from an observable and its analytical n-derivatives, and an “*a posteriori* error estimation” on each anharmonic force constant to control the accuracy of the least-squares fitting methods. In the framework of the (E-G) method consisting of fitting both *ab initio* electronic energies and gradients, we have used the experimental design approach in order to select two point distributions well suited for the construction of a complete quartic force field. In application to $H_2O$ and $H_2CO$ we compare the reduction of computational cost and variance obtained on each force constant with results from the standard least squares method.

## II. METHODOLOGY

### A. Standard least squares method

Let

$$V(s_1, s_2, ..., s_{nv}) = \sum_K^{Nterm} k_K X_K (s_1, s_2, ..., s_{nv}) \qquad (2)$$

be the reduced form of equation (1), where $k_K$ are the coefficients to be determined and $X_K$ the corresponding basis functions.

The best set of $k_K$ parameters is found by minimizing the following merit function (E method) :

$$R = \sum_m^{Npoint} \left[ \sum_K^{Nterm} k_K X_K (s_1, s_2, ..., s_{nv})_m - E_m \right]^2 \qquad (3)$$

It is worth noticing that this operation is equivalent to solving the following overdetermined system of linear equations :

$$\sum_K^{Nterm} k_K X_K (s_1, s_2, ..., s_{nv})_m \approx E_m \qquad m=1,\ldots, Npoint \qquad (4)$$

These equations can also be written in matrix notation as

$$[A] \times [k] = [B] \qquad (5)$$

where $[A]$ is an Npoint*Nterm matrix called “design matrix” with $A_{Km} = X_K(s_1, s_2, ..., s_{nv})_m$

and $[B]$ a vector of length Npoint such that $B_m = E_m$

Then, equation (5) is solved by writing :

$$[k] = [\alpha^{-1}]\,[\beta] \quad \text{with} \quad [\alpha^{-1}] = \{[A]^T [A]\}^{-1} \quad \text{and} \quad [\beta] = [A]^T [B] \tag{6}$$

B. Extended least squares method

Let $(\frac{\partial E}{\partial s_a})_m$ be the first derivative of the energy at the point m with respect to the $s_a$ coordinate and $\sum_{K=0}^{Nterm} k_K X'_{aK}(s_1, s_2, ..., s_{nv})_m$ the corresponding polynomial expansion. For each $s_a$ coordinate, there are Npoint equations :

$$\sum_{K=0}^{Nterm} k_K X'_{aK}(s_1, s_2, ..., s_{nv})_m \approx (\frac{\partial E}{\partial s_a})_m \tag{7}$$

Let $(\frac{\partial^2 E}{\partial s_a \partial s_b})_m$ be the second derivative of energy at the point m with respect to the $s_a$ and $s_b$ coordinates and $\sum_{K}^{Nterm} k_K X''_{abK}(s_1, s_2, ..., s_{nv})_m$ the corresponding polynomial expansion. For each pair of coordinates $s_a$ and $s_b$, we can also write Npoint equation :

$$\sum_{K}^{Nterm} k_K X''_{abK}(s_1, s_2, ..., s_{nv})_m \approx (\frac{\partial^2 E}{\partial s_a \partial s_b})_m \tag{8}$$

The three linear equations systems (4), (7), (8) which contain information about k parameters form a super overdetermined system :

$\sum_{K}^{Nterm} k_K X_K(s_1, s_2, ..., s_{nv})_m \approx E_m$ : *1 equation per point m*

$\sum_{K}^{Nterm} k_K X'_{aK}(s_1, s_2, ..., s_{nv})_m \approx (\frac{\partial E}{\partial s_a})_m$ : *nv equations per point m*

$$\sum_{K}^{Nterm} k_K X''_{ab\,K}(s_1, s_2, ..., s_{nv})_m \approx \left(\frac{\partial^2 E}{\partial s_a \partial s_b}\right)_m \; : \; \frac{1}{2}\textit{nv(nv+1) equations per point m}$$

Then this system is solved by application of the equation (6), *i.e* :

$$k_K = \sum_{j}^{Nterm} \alpha^{-1}_{Kj}\, \beta_j \tag{9}$$

with

$$\alpha_{Kj} = \sum_{m}^{Npoint} X_{Km} X_{mj} + \sum_{m}^{Npoint} \sum_{a}^{Nv} X'_{a\,Km} X'_{a\,mj} + \sum_{m}^{Npoint} \sum_{a}^{Nv} \sum_{b}^{Nv} X''_{abKm} X''_{ab\,mj}$$

and

$$\beta_j = \sum_{m}^{Npoint} X_{jm} E_m + \sum_{m}^{Npoint} \sum_{a}^{Nv} X'_{a\,jm} G_{a\,m} + \sum_{m}^{Npoint} \sum_{a}^{Nv} \sum_{b}^{Nv} X''_{ab\,jm} H_{ab\,m}$$

where $E_m$, $G_{am}$ and $H_{abm}$ are respectively energy, gradient and Hessian for each point k.

From the structure of the $[\alpha]$ and $[\beta]$ matrixes, it is very straightforward to obtain the expression of “extended least squares” method. When energies, gradients and Hessians are taken into account in one process (E-G-H method) the merit function take the following form :

$$R = \sum_{m=0.}^{Npoint} \left\{ \left[V(s_1, s_2, ..., s_{nv})_m - E_m\right]^2 + \left[\sum_{a}^{Nv} V'_a(s_1, s_2, ..., s_{nv})_m - \left(\frac{\partial E}{\partial s_a}\right)_m\right]^2 + \left[\sum_{a}^{Nv}\sum_{b}^{Nv} V''_{ab}(s_1, s_2, ..., s_{nv})_m - \left(\frac{\partial^2 E}{\partial s_a \partial s_b}\right)_m\right]^2 ... \right\}$$

*i.e*

$$R = R_E + R_G + R_H + ... \tag{10}$$

In this formalism, all available analytical data can jointly be used for obtaining all the polynomial coefficients. Thus, equation (10) ensures the best computational efficiency as well as the smoothing of numerical errors in comparison with finite difference formulas.

C. Error calculations

The variance on each k parameter to be determined may be estimated by

$$\sigma^2_{k_K} = \sum_m^{Npoint} \left(\frac{\partial k_K}{\partial E_m}\right)^2 \sigma^2_{E_l} + \sum_m^{Npoint} \sum_a^{Nv} \left(\frac{\partial k_K}{\partial G_{a_m}}\right)^2 \sigma^2_{G_{a_l}} + \sum_m^{Npoint} \sum_a^{Nv} \sum_b^{Nv} \left(\frac{\partial k_K}{\partial H_{ab_m}}\right)^2 \sigma^2_{H_{ab_m}}$$

according to equation (9), the above expression give[i] :

$$\sigma_{k_K} = \sqrt{\sum_j^{Nterm} \sum_l^{Nterm} \alpha^{-1}_{jK} \alpha^{-1}_{Kl} \left[ \sum_m^{Npoint} X_{jm} X_{lm} \sigma^2_{E_m} + \sum_m^{Npoint} \sum_a^{Nv} X'_{a_{jm}} X'_{a_{lm}} \sigma^2_{G_{a_m}} + \sum_m^{Npoint} \sum_a^{Nv} \sum_b^{Nv} X''_{ab_{jm}} X''_{ab_{lm}} \sigma^2_{H_{ab_m}} \right]} \quad (11)$$

Since the variances $\sigma^2_{E_m}$ , $\sigma^2_{G_{a_m}}$ , $\sigma^2_{H_{ab_m}}$ are unknown, they are estimated by the method of maximum likelihood[6] where the mean variance on energy ($\bar{\sigma}^2_E$), gradient ($\bar{\sigma}^2_G$) and Hessian ($\bar{\sigma}^2_H$) computation are related to the corresponding merit function :

$$\bar{s}^2_E = \frac{R_E}{Npoint}$$

$$\bar{s}^2_G = \frac{R_G}{Npoint * Nv}$$

$$\bar{s}^2_H = \frac{R_H}{\frac{1}{2} Npoint * Nv(Nv+1)}$$

The uncertainty on force constants $\sigma_{ijk}$, $\sigma_{ijkl}$ ( in $cm^{-1}$ ) is calculated from force constants expressed in internal coordinates by the relation :

---

[i] In the case of standard least squares method, equation (11) leads to the well known expression[6] :

$$\sigma^2_{k_K} = \alpha^{-1}_{KK} \ \bar{\sigma}^2_E$$

where the variance on each point of the grid ($\sigma^2_{E_l}$) has been approximated by the mean variance $\bar{\sigma}^2_E$.

$$\sigma_{\phi_{rst}} = \sqrt{\sum_{i,j,k}\left(\frac{\partial \phi_{rst}}{\partial f_{ijk}}\sigma_{f_{ijk}}\right)^2}$$

$$\sigma_{\phi_{rstu}} = \sqrt{\sum_{i,j,k,l}\left(\frac{\partial \phi_{rstu}}{\partial f_{ijkl}}\sigma_{f_{ijkl}}\right)^2}$$

with $\sigma_{f_K} = a!\,\sigma_{k_K}$ (cf. equation (1))

From the non-linear transformation of curvilinear coordinates into normal coordinates proposed by Hoy, Mills et Strey[7] and by only considering the linear terms of the transformation, we obtain an approximate expression of $\sigma_{\phi_{rst}}$ et $\sigma_{\phi_{rstu}}$ :

$$\sigma_{\phi_{rst}} = \sqrt{\sum_{i,j,k}\left(L_{ir}L_{js}L_{kt}\sigma_{f_{ijk}}\right)^2}$$

$$\sigma_{\phi_{rstu}} = \sqrt{\sum_{i,j,k,l}\left(L_{ir}L_{js}L_{kt}L_{lu}\sigma_{f_{ijkl}}\right)^2}$$

## III. EXAMPLE OF THE (E-G) method

In this case, the expression of the residue is truncated to first order derivatives: $R=R_E+R_G$

### A. Choice of the grid system

The grid system must fulfil three criteria:

- It must generate a non-singular $[\alpha] = [A]^T \cdot [A]$ matrix.
- It must be efficient, i.e. contain the minimum possible number of points for the determination of all parameters.
- It must provide sufficient accuracy for calculating coefficients, i.e. place the parameters within a confidence interval having little effect on the value of anharmonic vibrational frequencies.

This problem is far from being trivial and can be solved on the basis of work by Sana[8], who proposed several grids to determine a quadratic and cubic force field using a least square fitting of gradient data from the theory of experimental design. In this approach, an n-1 degree experimental plan is used to determine a complete n order force field. The simplex planes [9] thus lead to the determination of a harmonic force field with k+2 gradients for a non-symmetrical molecule with k variables. The composite planes[10] lead to the determination of a cubic force field with $1+2k+2^{k-p}$ gradients for k variables with p chosen to minimize the number of calculations[8].

Using the above procedure, a complete fourth order potential function can be determined with the (E-G) method by using a third degree design[11]. Two inexpensive grids were selected, the simplex-sum planes of Box and Behnken[12] truncated at the third sum and a Koshal[13] plan. The construction of these two planes is described in an appendix.

## B. Gain in the number of calculations

Table 1 shows savings in terms of calculation points of an (E-G) method for the two grids selected, in comparison to a standard least-square fitting procedure. Experience has shown that it is generally prudent to calculate at least twice as many points as coefficients when the usual linear regression method (E) is used, in order to smooth numerical errors. The use of the (E-G) method results in significant gains that increase with the size of the system studied, since the number of calculations is divided by a factor of about 5 for a triatomic molecule and close to 16 for a molecule containing 10 atoms. In addition, the redundancy of information with the (E-G) method also increases with the complexity of the problem, regardless of the plan used. If we compare the truncated simplex-sum design with the Koshal plan, it will be seen that the latter should in all cases be the most effective. Nevertheless it must be noted that molecular symmetry is not taken into account in this illustration.

## C. Gain in time

The computational cost of the analytical gradient is added to that for energy at each calculated point of the potential grid. Table 2 shows this increase with respect to the number of variables and to the *ab initio* method used. In this illustration, calculations were carried out at equilibrium geometry with the Gaussian 98 code[14]. For the HF/6-311G*, B3LYP/6-311G* and MP2/6-311G* methods, the mean increase in calculation times per point are 42 %, 29 % and 82 %, respectively It is thus low and constant, regardless of the complexity of the problem.

In fig. 1, the computational costs of the $H_2C_nO$ series at equilibrium geometry with the B3LYP/6-311G* method were multiplied by the number of points required to determine a quartic force field with the different linear regression methods. Thus, in spite of time limitations due to calculation of the gradient, the gain is satisfying since it rises by between a factor of 4 and 13 in the series shown and illustrates the degree of effectiveness of the (E-G) method. The accuracy of these grids is shown in the following examples.

## IV. CALCULATIONS OF QUARTIC FORCE FIELD WITH THE (E-G) METHOD

Using the examples of water and formaldehyde, we compared the non-null cubic and quartic terms calculated with the (E) and (E-G) methods, as well as the efficiency and accuracy of the two grids selected.

### A. Quartic force field of $H_2O$

44 calculations were needed with the type (E) regression method, and 11 and 10 with an (E-G) method, depending on the grid used in order to determine the 22 non-null parameters (see table 3).

In general and regardless of the plan used for the (E-G) method, the quartic force field was practically unchanged, since its mean difference is 0.6 $cm^{-1}$ for the simplex-sum grid and was 1.30 $cm^{-1}$ for the Koshal grid in comparison to the force field deduced with method (E) that we adopted as a reference.

All coefficients were well represented except for the term $k_{1133}$ determined with the (E-G) method using the Koshal grid which is not significant, since its value is twice its error bar. Whether this term is determined at -2.6 ± 0.4, -1.3 ± 0.5 or 0.0, its impact on the anharmonic correction of vibrational frequencies is minor[ii].

If the two plans are compared in terms of efficiency, taking the symmetry properties of the molecule into account, it is seen that the number of grid points that can be deduced[8] is 4 for the simplex-sum plan and 1 for the Koshal plan. Thus, the efficiency of the two plans is comparable in terms of the number of calculations.

---

[ii] i.e. a diagonal contribution (the greatest) of 1 $cm^{-1}$ on vibrational states $\nu_1$ and $\nu_3$.

In this example, it is nevertheless preferable to use the simplex-sum design since the mean error it generates on the force field, estimated at 0.3 $cm^{-1}$, is lower than that of the Koshal plan (0.9 $cm^{-1}$).

B. Quartic force field of $H_2CO$

The fourth order force field of formaldehyde (see table 4) has 84 non vanishing parameters. 168 calculations of the molecular wave function were carried out with the reference method, 44 with the (E-G) method using the simplex-sum plan, and 45 using the Koshal grid.

From a general standpoint, cubic force constants were correctly determined by the (E-G) method and the dispersion of mean error, of the order of 0.1 $cm^{-1}$, was low. Even so, the values of fourth order force constants are more uncertain, although the results converged within about 1 $cm^{-1}$. These non-significant values (in italic in the table) are in greater number when the (E-G) method is used.

Nevertheless, the 13 forces constants cancelled on the basis of the error criterion for the (E) method were also cancelled for the (E-G) method. In the case of the simplex-sum grid, the 6 additional parameters ($k_{1135}$, $k_{1144}$, $k_{2556}$, $k_{3345}$, $k_{3355}$, $k_{4456}$) cancelled by the (E-G) method but not by the (E) method are of only slight importance since they did not exceed 3 $cm^{-1}$. The Koshal grid poses more problems since it is evident that there are some relatively large terms poorly represented ($k_{2266}$, $k_{2666}$, $k_{3455}$, $k_{4555}$, $k_{4455}$ ).

Comparison of the two plans used for the (E-G) method shows that the simplex-sum grid is more accurate since it leads to a lower mean error, lower dispersion, and results that are in better agreement with the reference method. It is also just as effective as the Koshal plan since it enables better use of the symmetry properties of the molecule.

## V. CONCLUSION

The REGRESS EGH code developed in our laboratory enables the inclusion of energies and their $n^{th}$ analytical derivative with respect to nuclear parameters in the same linear regression process. This leads to the determination of a complete fourth order force field by optimal use of available data on the molecular wave function. We have also implemented an algorithm to calculate the statistical error of polynomial parameters. This approach enables the selection of significant parameters in the description of the potential.

Using the example of complete quartic force fields of the water and formaldehyde molecules, we have illustrated the accuracy obtained with the (E-G) method with two grids. The primary conclusion from this work is that the simplex-sum grid truncated at the third sum provides results in excellent agreement with results obtained with the type (E) method. In terms of efficiency, it is even better than the Koshal plan when the molecule has some symmetry.

This code tested on small size systems, should be useful for the anharmonic vibrational treatment of bigger size molecules. To this end, the vibrational study of acetonitrile[15] and its efficiency of treatment due to the (E-G) method will be presented in a next paper.

Acknowledgements

One of the authors (Ph. C.) thanks the Conseil Regional d'Aquitaine for a grant. We acknowledge the Centre Informatique National de l'Enseignement Superieur (CINES) for support of this work.. We express our sincere gratitude to Pr. Ross Brown for helpful discussions.

# APPENDIX : CONSTRUCTION OF SIMPLEX-SUM AND KOSCHAL DESIGNS

## (example with four variables)

### Simplex sum design

| $Npoint_1$ | Nv 1 | 2 | 3 | 4 | … | k |
|---|---|---|---|---|---|---|
| [1] | $-1/2$ | $-1/2\sqrt{3}$ | $-1/2\sqrt{6}$ | $-1/2\sqrt{10}$ | … | $-1/\sqrt{2k(k+1)}$ |
| [2] | $1/2$ | $-1/2\sqrt{3}$ | $-1/2\sqrt{6}$ | $-1/2\sqrt{10}$ | … | $-1/\sqrt{2k(k+1)}$ |
| [3] | 0 | $2/2\sqrt{3}$ | $-1/2\sqrt{6}$ | $-1/2\sqrt{10}$ $\mathbf{a_1}$ | … | $-1/\sqrt{2k(k+1)}$ |
| [4] | 0 | 0 | $3/2\sqrt{6}$ | $-1/2\sqrt{10}$ | … | $-1/\sqrt{2k(k+1)}$ |
| [5] | 0 | 0 | 0 | $4/2\sqrt{10}$ | … | $-1/\sqrt{2k(k+1)}$ |
| … | … | … | … | … | … | … |
| [Nv+1] | 0 | 0 | 0 | 0 | … | $k/\sqrt{2k(k+1)}$ |

$a_1$ matrix or first sum matrix : Simplex grid

| $Npoint_2$ | Nv 1 | 2 | 3 | 4 |
|---|---|---|---|---|
| [1]+[2] | 0 | $-2/2\sqrt{3}$ | $-2/2\sqrt{6}$ | $-2/2\sqrt{10}$ |
| [1]+[3] | $-1/2$ | $1/2\sqrt{3}$ | $-2/2\sqrt{6}$ | $-2/2\sqrt{10}$ |
| [1]+[4] | $-1/2$ | $-1/2\sqrt{3}$ | $2/2\sqrt{6}$ | $-2/2\sqrt{10}$ $\mathbf{a_2}$ |
| … | … | … | … | … |
| [i]+[j<i] | | | | |

$a_2$ matrix or second sum matrix

| $Npoint_3$ | Nv 1 | 2 | 3 | 4 |
|---|---|---|---|---|
| [1]+[2]+[3] | 0 | 0 | $-3/2\sqrt{6}$ | $-3/2\sqrt{10}$ |
| [1]+[2]+[4] | 0 | $-2/2\sqrt{3}$ | $1/2\sqrt{6}$ | $-3/2\sqrt{10}$ |
| [1]+[2]+[5] | 0 | $-2/2\sqrt{3}$ | $-2/2\sqrt{6}$ | $2/2\sqrt{10}$ $\mathbf{a_3}$ |
| … | … | … | … | … |
| [i]+[j>i]+[k>j] | | | | |

$a_3$ matrix or third sum matrix

| $Npoint_4$ | Nv 1 | 2 | 3 | 4 |
|---|---|---|---|---|
| [1]+[2]+[3]+[4] | 0 | 0 | 0 | $-4/2\sqrt{10}$ |
| [1]+[2]+[3]+[5] | 0 | 0 | $-3/2\sqrt{6}$ | $1/2\sqrt{10}$ |
| [1]+[2]+[4]+[5] | 0 | $-2/2\sqrt{3}$ | $1/2\sqrt{6}$ | $1/2\sqrt{10}$ $\mathbf{a_4}$ |
| … | … | … | … | … |
| [i]+[j>i]+[k>j]+ [l>k] | | | | |

$a_4$ matrix or fourth sum matrix

| $Npoint_5$ | Nv 1 | 2 | 3 | 4 |
|---|---|---|---|---|
| [1]+[2]+[3]+[4]+[5] | 0 | 0 | 0 | 0 |

$a_5$ matrix or fifth sum matrix, reference geometry.

### Koshal type design

| Npoint | Nv 1 | 2 | 3 | 4 |
|---|---|---|---|---|
| [1] | 0 | 0 | 0 | 0 |
| Reference geometry | | | | |
| [2] | 1 | 0 | 0 | 0 |
| [3] | 0 | 1 | 0 | 0 |
| [4] | 0 | 0 | 1 | 0 |
| [5] | 0 | 0 | 0 | 1 |
| [6] | -1 | 0 | 0 | 0 |
| [7] | 0 | -1 | 0 | 0 |
| [8] | 0 | 0 | -1 | 0 |
| [9] | 0 | 0 | 0 | -1 |
| Simple steps | | | | |
| [10] | -1 | -1 | 0 | 0 |
| [11] | -1 | 0 | -1 | 0 |
| [12] | -1 | 0 | 0 | -1 |
| [13] | 0 | -1 | -1 | 0 |
| [14] | 0 | -1 | 0 | -1 |
| [15] | 0 | 0 | -1 | -1 |
| Double steps | | | | |
| [16] | 1 | 1 | 1 | 0 |
| [17] | 1 | 1 | 0 | 1 |
| [18] | 1 | 0 | 1 | 1 |
| [19] | 0 | 1 | 1 | 1 |
| Triple steps | | | | |

Note :

*designs are built up in the internal factor space

*$\alpha_i$ are scale factors. In our work,

$\alpha_1 = \sqrt{2k(k+1)}/k$ , $\alpha_2 = 2\sqrt{2k(k+1)}/k$ ,

$\alpha_3 = 3\sqrt{2k(k+1)}/k$ for the truncated simplex sum design to the third sum

Table 1 : Number of calculations with the (E-G) method for the two grids selected. Comparison with a usual (E) linear regression method (without symmetry consideration).

| Nv | Nterm | (E) | | Simplex-sum design (E-G) | | | Koshal design(E-G) | | |
|---|---|---|---|---|---|---|---|---|---|
| | | Npoint | Redundancy | Npoint | Redundancy | Gain | Npoint | Redundancy | Gain |
| 3 | 35 | 70 | 2 | 15 | 1.7 | 4.7 | 11 | 1.3 | 6.4 |
| 6 | 210 | 420 | 2 | 64 | 2.1 | 6.6 | 48 | 1.6 | 8.8 |
| 9 | 715 | 1430 | 2 | 176 | 2.5 | 8.1 | 139 | 1.9 | 10.3 |
| 12 | 1820 | 3640 | 2 | 378 | 2.7 | 9.6 | 311 | 2.2 | 11.7 |
| 15 | 3876 | 7752 | 2 | 697 | 2.9 | 11.1 | 591 | 2.4 | 13.1 |
| 18 | 7315 | 14630 | 2 | 1160 | 3.0 | 12.6 | 1006 | 2.6 | 14.5 |
| 21 | 12650 | 25300 | 2 | 1794 | 3.1 | 14.1 | 1583 | 2.8 | 16.0 |
| 24 | 20475 | 40950 | 2 | 2626 | 3.2 | 15.6 | 2349 | 2.9 | 17.4 |

Table 2 : Time limitation due to calculation of the gradient according to different level of theory.

| | | $\frac{t(E+G)}{t(E)}$ | | |
|---|---|---|---|---|
| Molecule ($C_2V$) | Nv | HF/6-311G* | B3LYP/6-311G* | MP2/6-311G* |
| $H_2O$ | 3 | 1.45 | 1.44 | 1.76 |
| $H_2CO$ | 6 | 1.47 | 1.31 | 1.7 |
| $H_2C_2O$ | 9 | 1.59 | 1.27 | 1.71 |
| $H_2C_3O$ | 12 | 1.51 | 1.25 | 1.87 |
| $H_2C_4O$ | 15 | 1.34 | 1.27 | 1.79 |
| $H_2C_5O$ | 18 | 1.34 | 1.33 | 2.08 |
| $H_2C_6O$ | 21 | 1.36 | 1.22 | 1.79 |
| $H_2C_7O$ | 24 | 1.30 | 1.22 | 1.83 |

Calculations were carried out with the Gaussian 98 package on an HP 700 MHz/1GO RAM work station.

Tableau 3 : Quartic force field of water calculated at B3LYP/cc-pVTZ level of theory

| $H_2O$ (Nterm=22) | | | Regression (E) | Regression (E-G) : simplex-sum grid | Regression (E-G) : Koshal grid |
|---|---|---|---|---|---|
| Actual number of point[a] | | | 44 | 11 | 10 |
| Seconde derivatives ($cm^{-1}$) | | | | | |
| $\omega_1$ | $\delta_{HOH}$ | $A_1$ | 1639.13 | 1639.11 | 1639.33 |
| $\omega_2$ | $\nu_{sym\ OH}$ | $A_1$ | 3799.52 | 3799.56 | 3799.56 |
| $\omega_3$ | $\nu_{asym\ OH}$ | $B_1$ | 3899.80 | 3899.83 | 3899.77 |
| Cubic parameters[b] ($cm^{-1}$) | | | | | |
| $k_{111}$ | | | -58.68 ± 0.07 | -58.33 ± 0.05 | -58.53 ± 0.02 |
| $k_{112}$ | | | -46.38 ± 0.08 | -46.00 ± 0.10 | -46.97 ± 0.07 |
| $k_{221}$ | | | 60.20 ± 0.10 | 60.10 ± 0.10 | 59.12 ± 0.08 |
| $k_{222}$ | | | -301.30 ± 0.20 | -301.80 ± 0.10 | -301.55 ± 0.04 |
| $k_{331}$ | | | 116.46 ± 0.07 | 115.90 ± 0.10 | 116.00 ± 0.10 |
| $k_{332}$ | | | -906.72 ± 0.09 | -906.40 ± 0.20 | -907.90 ± 0.10 |
| Quartic parameters[b] ($cm^{-1}$) | | | | | |
| $k_{1111}$ | | | -4.3 ± 0.4 | -4.8 ± 0.1 | -4.8 ± 0.05 |
| $k_{1112}$ | | | 7.8 ± 0.8 | 7.3 ± 0.3 | 8.1 ± 0.4 |
| $k_{1122}$ | | | 3.3 ± 0.6 | 4.3 ± 0.5 | 5.0 ± 2.0 |
| $k_{1133}$ | | | -2.6 ± 0.4 | -1.3 ± 0.5 | *(0.0)* *-3.0 ± 2.0* |
| $k_{2221}$ | | | -14.0 ± 2.0 | -13.8 ± 0.6 | -12.6 ± 0.7 |
| $k_{2222}$ | | | 31.0 ± 1.0 | 31.6 ± 0.4 | 31.0 ± 0.1 |
| $k_{2233}$ | | | 192.7 ± 0.6 | 191.0 ± 0.8 | 189.0 ± 3.0 |
| $k_{3312}$ | | | -45.8 ± 0.8 | -45.1 ± 0.9 | -41.0 ± 4.0 |
| $k_{3333}$ | | | 32.3 ± 0.4 | 32.0 ± 0.2 | 31.5 ± 0.2 |
| RMS error ($cm^{-1}$) | | | 0.0027 | 0.0032 | 0.0032 |
| Mean $\left\| k_{k(E-G)} - k_{k(E)} \right\|$ ($cm^{-1}$) | | | 0 | 0.6 | 1.3 |
| Mean $\left\| \sigma k_k \right\|$ ($cm^{-1}$) | | | 0.5 | 0.3 | 0.9 |

Interpolation domain : $\Delta R_{OH}$=0.03 A, $\Delta\alpha$=7.5°
Stationary point : $R_{OH}$=0.9613 A, $\alpha$=104.52°
[a]effective point number to be computed taking into account the symmetry reduction (see Ref. 7)
[b]Anharmonic constants are given with their corresponding error : $k_k \pm \sigma k_k$. $k_k$ are cancelled if $k_k < 2\sigma k_k$.

Tableau 4 : Quartic force field of formaldehyde calculated at B3LYP/cc-pVTZ level of theory

| $H_2CO$ (Nterm=84) | | | Regression (E) | Regression (E-G) : simplex-sum grid | Regression (E-G) : Koshal grid |
|---|---|---|---|---|---|
| Actual number of point[a] | | | 168 | 44 | 45 |
| Seconde derivatives ($cm^{-1}$) | | | | | |
| $\omega_1$ | $\gamma_{CH2}$ | $B_2$ | 1202.31 | 1203.68 | 1201.86 |
| $\omega_2$ | $rock_{CH}$ | $B_1$ | 1267.91 | 1267.87 | 1267.87 |
| $\omega_3$ | $\delta_{CH2}$ | $A_1$ | 1536.03 | 1536.21 | 1536.04 |
| $\omega_4$ | $\nu_{CO}$ | $A_1$ | 1823.05 | 1822.62 | 1823.16 |
| $\omega_5$ | $\nu_{sym\ CH}$ | $A_1$ | 2876.45 | 2875.93 | 2876.54 |
| $\omega_6$ | $\nu_{asym\ CH}$ | $B_1$ | 2930.65 | 2929.61 | 2929.92 |
| Cubic parameters[b] ($cm^{-1}$) | | | | | |
| $k_{113}$ | | | 58.82 ± 0.06 | 57.00 ± 0.20 | 57.90 ± 0.30 |
| $k_{114}$ | | | 39.90 ± 0.10 | 39.10 ± 0.30 | 38.90 ± 0.50 |
| $k_{115}$ | | | -43.10 ± 0.20 | -42.30 ± 0.40 | -41.43 ± 0.70 |
| $k_{223}$ | | | -72.61 ± 0.02 | -73.23 ± 0.06 | -73.03 ± 0.10 |
| $k_{224}$ | | | 20.88 ± 0.04 | 20.60 ± 0.10 | 20.40 ± 0.20 |
| $k_{225}$ | | | -13.87 ± 0.07 | -13.60 ± 0.20 | -13.05 ± 0.30 |
| $k_{236}$ | | | 17.39 ± 0.09 | 17.40 ± 0.10 | 17.16 ± 0.40 |
| $k_{246}$ | | | 28.40 ± 0.20 | 28.70 ± 0.30 | 30.10 ± 0.80 |
| $k_{256}$ | | | -5.20 ± 0.30 | -5.80 ± 0.40 | -7.00 ± 1.00 |
| $k_{333}$ | | | 10.39 ± 0.02 | 10.46 ± 0.02 | 10.48 ± 0.04 |
| $k_{334}$ | | | 70.60 ± 0.05 | 71.20 ± 0.08 | 71.40 ± 0.20 |
| $k_{335}$ | | | -26.16 ± 0.08 | -26.00 ± 0.10 | -26.00 ± 0.20 |
| $k_{344}$ | | | -44.10 ± 0.10 | -43.90 ± 0.20 | -43.20 ± 0.30 |
| $k_{345}$ | | | -20.50 ± 0.40 | -19.70 ± 0.30 | -20.20 ± 0.70 |
| $k_{355}$ | | | -13.10 ± 0.20 | -13.40 ± 0.30 | -12.70 ± 0.70 |
| $k_{366}$ | | | -58.60 ± 0.10 | -58.50 ± 0.30 | -59.50 ± 0.60 |
| $k_{444}$ | | | 98.70 ± 0.10 | 98.40 ± 0.20 | 99.00 ± 0.30 |
| $k_{445}$ | | | 9.00 ± 0.20 | 8.10 ± 0.40 | 9.20 ± 0.80 |
| $k_{455}$ | | | -21.50 ± 0.30 | -22.50 ± 0.60 | -23.00 ± 1.00 |
| $k_{466}$ | | | -61.20 ± 0.20 | -62.10 ± 0.60 | -63.00 ± 1.00 |
| $k_{555}$ | | | -223.00 ± 0.50 | -225.00 ± 0.60 | -225.00 ± 1.00 |
| $k_{566}$ | | | -711.00 ± 0.30 | -710.60 ± 0.90 | -711.00 ± 2.00 |
| Quartic parameters[b] ($cm^{-1}$) | | | | | |
| $k_{1111}$ | | | 6.70 ± 0.10 | 6.50 ± 0.50 | 7.40 ± 0.90 |
| $k_{1122}$ | | | -9.65 ± 0.06 | -10.10 ± 0.30 | -11.30 ± 0.40 |
| $k_{1126}$ | | | 5.90 ± 0.20 | 6.40 ± 0.60 | 10.00 ± 2.00 |
| $k_{1133}$ | | | -3.16 ± 0.06 | -3.70 ± 0.20 | -3.50 ± 0.20 |
| $k_{1134}$ | | | -1.60 ± 0.20 | -2.40 ± 0.40 | -3.40 ± 0.80 |

| | | | | | | |
|---|---|---|---|---|---|---|
| $k_{1135}$ | | 1.10 ± 0.30 | *(0.0)* | *0.70 ± 0.50* | *(0.0)* | *2.00 ± 1.00* |
| $k_{1144}$ | | -1.00 ± 0.20 | *(0.0)* | *-1.00 ± 1.00* | *(0.0)* | *-1.80 ± 0.90* |
| $k_{1145}$ | *(0.0)* | *-0.90 ± 0.60* | *(0.0)* | *0.00 ± 1.00* | *(0.0)* | *4.00 ± 2.00* |
| $k_{1155}$ | *(0.0)* | *0.90 ± 0.50* | *(0.0)* | *2.00 ± 1.00* | *(0.0)* | *2.00 ± 2.00* |
| $k_{1166}$ | | -7.30 ± 0.30 | | -7.00 ± 1.00 | | -9.00 ± 2.00 |
| $k_{2222}$ | | 1.73 ± 0.02 | | 1.80 ± 0.20 | | 1.80 ± 0.20 |
| $k_{2226}$ | *(0.0)* | *0.00 ± 2.00* | *(0.0)* | *0.10 ± 0.30* | *(0.0)* | *-1.00 ± 1.00* |
| $k_{2233}$ | | 1.00 ± 0.02 | | 1.27 ± 0.06 | | 0.90 ± 0.10 |
| $k_{2234}$ | | -5.28 ± 0.08 | | -5.90 ± 0.20 | | -6.20 ± 0.30 |
| $k_{2235}$ | | 3.00 ± 0.10 | | 2.80 ± 0.20 | | 3.10 ± 0.50 |
| $k_{2244}$ | | -1.36 ± 0.08 | | -1.10 ± 0.20 | | -1.90 ± 0.30 |
| $k_{2245}$ | | -3.10 ± 0.20 | | -3.20 ± 0.40 | *(0.0)* | *-0.70 ± 0.90* |
| $k_{2255}$ | | -2.40 ± 0.20 | | -1.90 ± 0.50 | | -2.00 ± 0.80 |
| $k_{2266}$ | *(0.0)* | *0.00 ± 8.00* | *(0.0)* | *-2.00 ± 1.00* | *(0.0)* | *4.00 ± 4.00* |
| $k_{2336}$ | | 1.47 ± 0.09 | | 1.97 ± 0.10 | | 3.00 ± 0.30 |
| $k_{2346}$ | | 1.80 ± 0.30 | | 2.30 ± 0.30 | *(0.0)* | *2.00 ± 1.00* |
| $k_{2356}$ | | -3.60 ± 0.50 | | -2.50 ± 0.40 | *(0.0)* | *-2.00 ± 2.00* |
| $k_{2446}$ | | 3.80 ± 0.30 | | 3.20 ± 0.40 | | 5.00 ± 1.00 |
| $k_{2456}$ | *(0.0)* | *0.60 ± 0.90* | *(0.0)* | *1.00 ± 1.00* | *(0.0)* | *-5.00 ± 3.00* |
| $k_{2556}$ | | 2.80 ± 0.70 | *(0.0)* | *2.00 ± 1.00* | *(0.0)* | *4.00 ± 3.00* |
| $k_{2666}$ | *(0.0)* | *-1.00 ± 9.00* | *(0.0)* | *2.00 ± 1.00* | *(0.0)* | *-8.00 ± 5.00* |
| $k_{3333}$ | | 1.61 ± 0.02 | | 1.58 ± 0.02 | | 1.59 ± 0.06 |
| $k_{3334}$ | | 1.80 ± 0.10 | | 1.55 ± 0.10 | | 1.65 ± 0.40 |
| $k_{3335}$ | *(0.0)* | *0.10 ± 0.20* | *(0.0)* | *-0.20 ± 0.20* | *(0.0)* | *-0.40 ± 0.40* |
| $k_{3344}$ | | 4.80 ± 0.20 | | 4.70 ± 0.30 | | 5.00 ± 1.00 |
| $k_{3345}$ | | -1.30 ± 0.40 | *(0.0)* | *-1.00 ± 0.60* | *(0.0)* | *-2.00 ± 1.00* |
| $k_{3355}$ | | -0.60 ± 0.20 | *(0.0)* | *0.00 ± 0.50* | *(0.0)* | *1.00 ± 2.00* |
| $k_{3366}$ | | -6.60 ± 0.10 | | -6.10 ± 0.40 | | -7.20 ± 0.50 |
| $k_{3444}$ | | -5.40 ± 0.40 | | -4.70 ± 0.30 | | -4.00 ± 1.00 |
| $k_{3445}$ | *(0.0)* | *-1.20 ± 0.90* | *(0.0)* | *-1.50 ± 0.80* | *(0.0)* | *4.00 ± 3.00* |
| $k_{3455}$ | *(0.0)* | *0.70 ± 0.70* | *(0.0)* | *2.00 ± 1.00* | *(0.0)* | *7.00 ± 4.00* |
| $k_{3466}$ | | -6.50 ± 0.50 | | -6.20 ± 0.90 | | -8.00 ± 2.00 |
| $k_{3555}$ | *(0.0)* | *2.00 ± 2.00* | *(0.0)* | *4.00 ± 2.00* | *(0.0)* | *3.00 ± 4.00* |
| $k_{3566}$ | | 17.00 ± 0.70 | | 17.00 ± 1.00 | | 17.00 ± 2.00 |
| $k_{4444}$ | | 7.00 ± 0.30 | | 6.80 ± 0.40 | | 8.30 ± 0.70 |
| $k_{4445}$ | *(0.0)* | *2.40 ± 1.20* | *(0.0)* | *1.70 ± 0.90* | *(0.0)* | *-2.00 ± 3.00* |
| $k_{4455}$ | *(0.0)* | *0.90 ± 0.70* | *(0.0)* | *1.00 ± 2.00* | *(0.0)* | *-5.00 ± 5.00* |
| $k_{4466}$ | | -3.30 ± 0.40 | *(0.0)* | *-1.00 ± 1.00* | *(0.0)* | *-4.00 ± 2.00* |
| $k_{4555}$ | *(0.0)* | *3.10 ± 2.90* | *(0.0)* | *4.00 ± 3.00* | *(0.0)* | *-13.00 ± 7.00* |
| $k_{4566}$ | | 16.00 ± 1.00 | | 17.00 ± 2.00 | | 22.00 ± 4.00 |
| $k_{5555}$ | | 22.00 ± 2.00 | | 21.00 ± 2.00 | | 21.00 ± 4.00 |

| | | | |
|---|---|---|---|
| $k_{5566}$ | 145.00 ± 1.00 | 146.00 ± 3.00 | 142.00 ± 4.00 |
| $k_{6666}$ | 25.10 ± 0.70 | 25.00 ± 3.00 | 30.00 ± 4.00 |
| RMS error ($cm^{-1}$) | 0.05 | 0.16 | 0.26 |
| Mean $\left\|k_{k(E-G)} - k_{k(E)}\right\|$ ($cm^{-1}$) | 0 | 0.5 | 1.0 |
| Mean $\left\|\sigma k_k\right\|$ ($cm^{-1}$) | 0.7 | 0.6 | 1.5 |

Interpolation domain : $\Delta R_{CO}$ =0.07 A, $\Delta R_{CH}$ =0.07 A, $\Delta\alpha_{OCH}$=14°, $\Delta\beta_{H1COH2}$=21°
Stationary point : $R_{CO}$=1.1990 A, $R_{CH}$=1.1068 A, $\alpha_{OCH}$=122.14°, $\beta_{H1COH2}$=180°
[a]actual number of point to be computed taking into account the symmetry reduction (see Ref. 7)
[b]Anharmonic constants are given with their corresponding error : $k_k \pm \sigma k_k$. $k_k$ are cancelled if $k_k < 2\sigma k_k$.

CAPTIONS

Fig. 1 : Computation cost (CPU time) of a quartic force field with the different linear regression methods (E) and (E-G with simplex-sum grid) for the $H_2C_nO$ series.

Fig. 1 : Computation cost (CPU time) of a quartic force field with the different linear regression methods (E) and (E-G with simplex-sum grid) for the $H_2C_nO$ series.

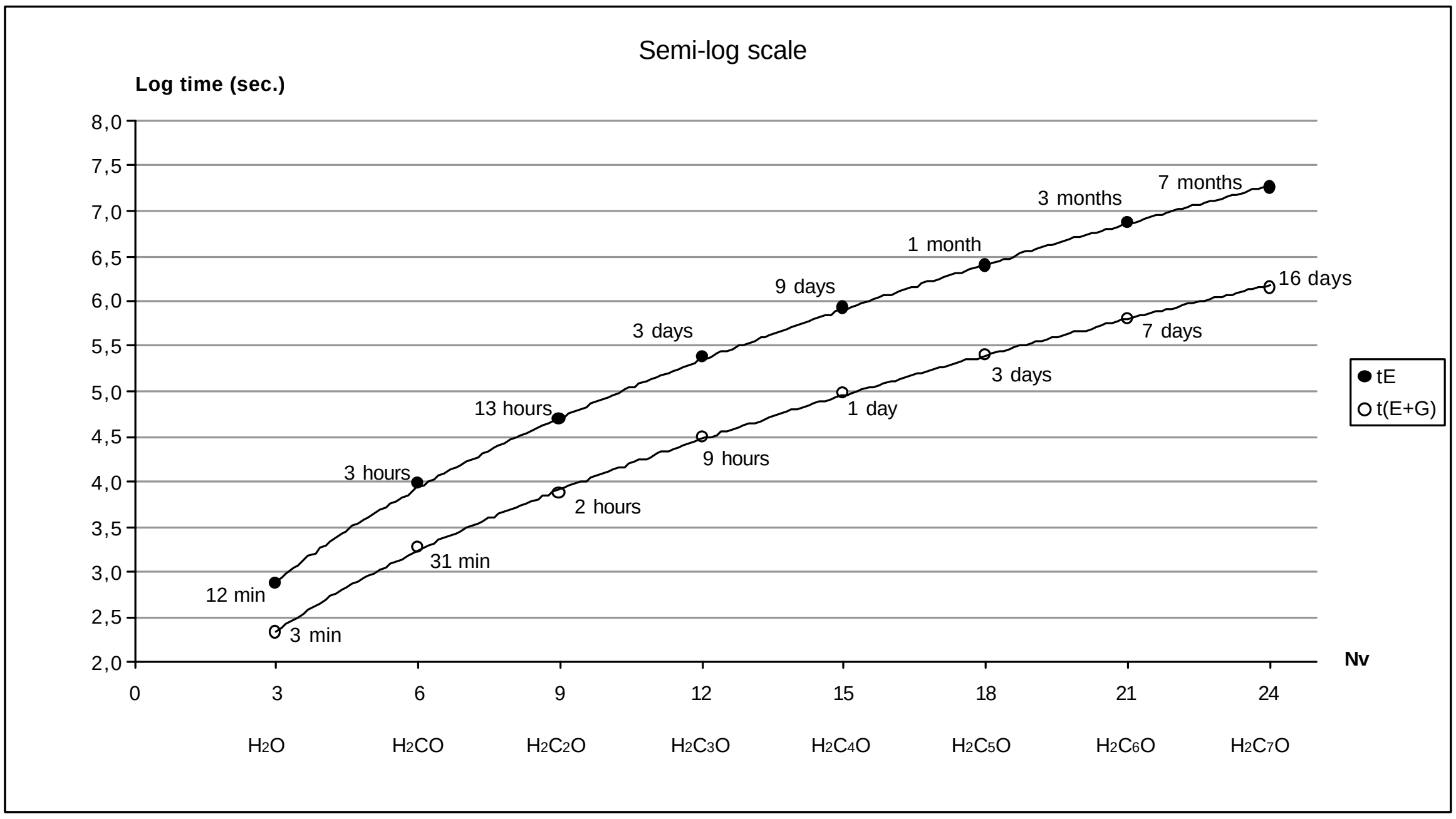

Calculations were carried out with the B3LYP/6-311G* method in the Gaussian 98 package on an HP 700 MHtz/ 1GO. RAM workstation (without symmetry consideration).